\documentclass[reprint,superscriptaddress,showpacs,preprintnumbers,nofootinbib,amsmath,amssymb,aps,onecolumn,prd,
floatfix
]{revtex4}

\usepackage{graphicx}
\usepackage{epsfig}
\usepackage{dcolumn}
\usepackage{bm}
\usepackage{amssymb}
\usepackage{amsmath}
\usepackage{subfigure}
\usepackage{cancel}
\usepackage[colorlinks]{hyperref}
\usepackage[usenames,dvipsnames]{color}

\hypersetup{
     breaklinks=true,
    pdfstartview={FitH},    
    colorlinks=true,       
    linkcolor=blue,          
    citecolor=red,        
    filecolor=magenta,      
    urlcolor=blue,           
    anchorcolor=green,      
    linktocpage=true
}
\newcommand{\Mpl}{M_{\textrm{Pl}}}

\newcommand{\nn}{\nonumber}

\def\lam{\lambda}

\def\S{\mathcal{S}}

\def\doi{http://doi.org}
\def\arxiv{http://arxiv.org/abs}

\def\r{\mathrm{r}}

\def\m{\mathrm{m}}

\def\d{\mathrm{d}}

\def\doi{http://doi.org}
\def\arxiv{http://arxiv.org/abs}

\begin{document}

\title{Can massive neutrinos be responsible for late time phase transition $\hat{\rm a}$ {\it  la} deceleration to acceleration in our Universe?}

\author{M.~Sami}
\email{samijamia@gmail.com}  \affiliation{International Center for Cosmology, Charusat University, Anand 388421, Gujarat, India}\affiliation{Institute for Advanced Physics $\&$ Mathematics, Zhejiang University of Technology,\\ Hangzhou, 310032, China} \affiliation{Center for Theoretical Physics, Eurasian National University, Astana 010008, Kazakhstan}	
\author{Shynaray Myrzakul}
\email{srmyrzakul@gmail.com}
\affiliation{Center for Theoretical Physics, Eurasian National University, Astana 010008, Kazakhstan}	
\author{Mudhahir Al Ajmi} \email{mudhahir@gmail.com}
\affiliation{ Department of Physics, College of Science, Sultan Qaboos University, P.O. Box 36, Al-Khodh 123, Muscat, Sultanate of Oman}

\begin{abstract}
We attempt a novel mechanism to understand the underlying cause of
late-time cosmic acceleration using a distinguished physical process
taking place in the late Universe. The turning of massive neutrinos from relativistic to
non-relativistic might cause a phase transition at late times. We 
implement this idea using massless $\lambda \phi^4$ theory coupled to
massive neutrino matter such that the coupling is proportional to
the trace of the energy-momentum tensor of neutrino matter. As massive
neutrinos become non-relativistic, their coupling to the scalar field builds up dynamically
giving rise to spontaneous symmetry breaking in the low-density regime. 
As a result, in the true
vacuum, the field acquires non zero mass proportional to the energy
density of massive neutrino matter and could give rise to late-time
cosmic acceleration.  We also address the issues related to stability of self coupling under radiative corrections.
\end{abstract}

\maketitle

 The observed transition from deceleration to acceleration, Universe has recently undergone through, is termed as one of the most remarkable discoveries of modern cosmology \cite{Sahni, Copeland} though the underlying cause of the phenomenon yet remains to be unveiled. It is plausible to think if there is a distinguished physical process in the Universe that could be responsible for the late time transition. One of the important physical processes in the Universe is associated with massive neutrinos \cite{Les} which turn non-relativistic at late stages. It is mysterious that mass scale, in this case, is around the one associated with dark energy. It is intriguing to ask whether massive neutrinos could trigger the transition to late time acceleration?
 
\begin{figure}[ht]
 \centering
\includegraphics[scale=.4]{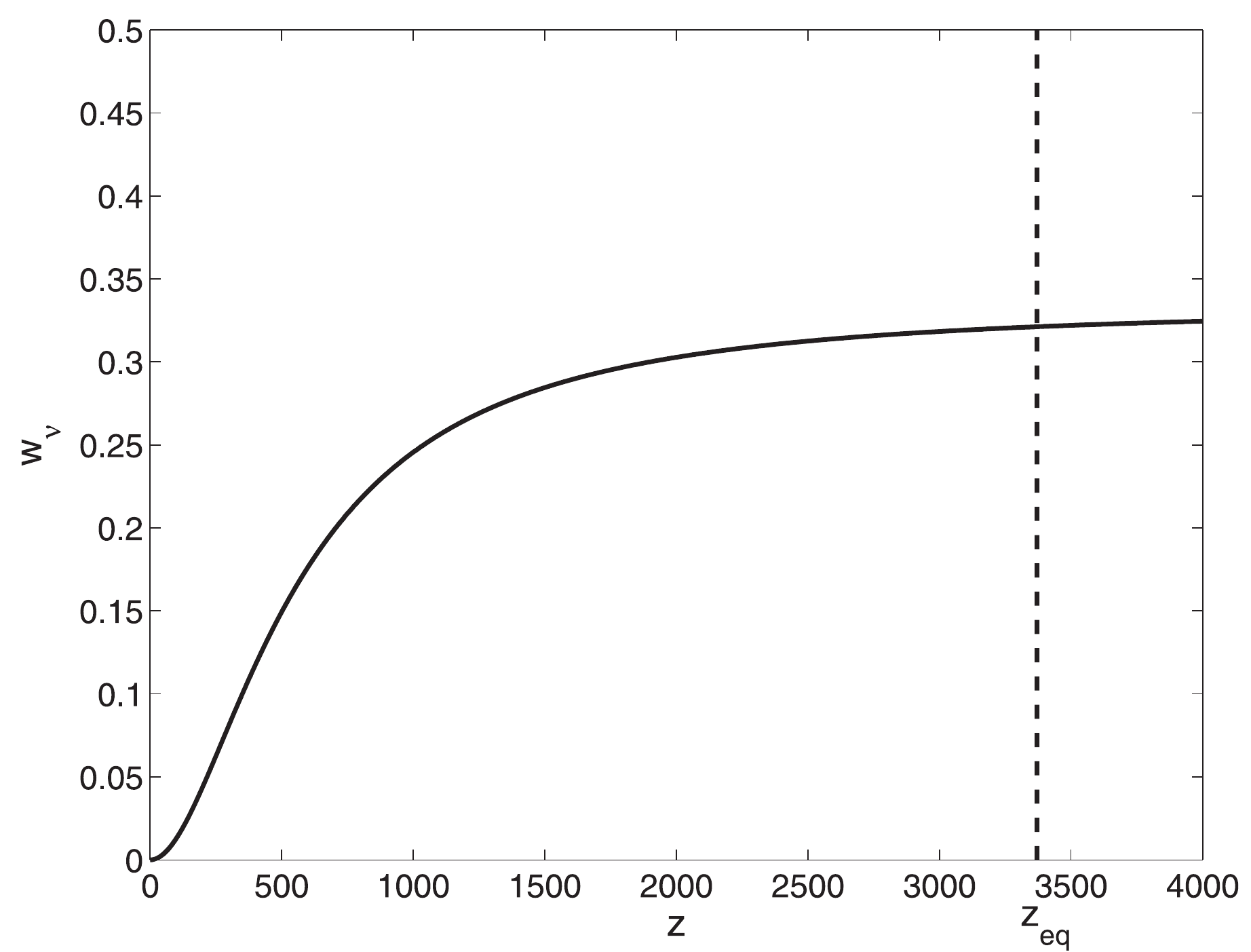}
\caption{The figure shows the equation of state parameter $\omega_\nu$ versus $N\equiv\ln a$ in the framework of $\Lambda$CDM. The figure shows how $\omega_\nu$ changes from $1/3$(radiation like) to $\omega_\nu=0$(cold matter like) as Universe evolves and massive neutrinos turn from relativistic to non-relativistic. Phase transition commences  as $\omega_\nu$ starts deviating from $1/3$ and completes when it settles zero which happens about the present epoch.} \label{wnu}
\end{figure}

 The idea of late-time phase transition $\hat{\rm a}$  {\it  la} symmetron was first attempted  in Ref~\cite{Hinterbichler:2010es}, see also Refs~\cite{Bamba:2012yf,cop,others1,others3,Amend,Ant,novikov} on the related theme\footnote{The proposal for direct coupling of matter with dark energy(quintessence) was made by Amendola in his pioneering paper\cite{Amend}, similar concept was applied later to massive neutrino matter in Refs\cite{Amendola,Wet}}. In that case, the $Z_2$ symmetry was exact locally in high-density regime; symmetry was broken at large scales as matter density in the Universe approached the critical density. Unfortunately, the idea did not succeed as the local gravity constraints forced the mass of the scalar field to assume numerical values much higher than $H_0$~\cite{ Hinterbichler:2010es}. Consequently, despite, the local minimum in the field potential, the field could not settle there due the absence of slow roll around the minimum thereby the model failed to account for late-time cosmic acceleration. An interesting attempt was made in Ref.~\cite{cop} to realize symmetry breaking in low-density regime using radiating correction $\hat{\rm a}$  {\it  la} Coleman-Weinberg.\\
 It is quite tempting to bring in the massive neutrino matter thank to the aforesaid features. In this case, we would not conflict with local physics. Secondly, at present, there is no convincing support of dark energy interaction with cold dark(baryonic) matter though a similar question remains to be an open one for massive neutrino matter \cite{Amendola,Wet,staro, H1}. Needless to mention that proper local screening of chameleon/symmetron leaves no scope for self acceleration in these scenarios \cite{khoury}.\\ 
Motivated by the above argument, let us consider the massless $\lambda \phi^4$ theory coupled to massive neutrino matter such that the coupling is proportional to its trace, $T_\nu=(3\omega_\nu-1)$. Such a coupling builds up only at late stages as massive neutrinos turn non-relativistic, see Fig.\ref{wnu}. At early times, the coupling is absent thereby, $\phi=0$ is minimum in the theory. In case, tachyonic instability occurs in the low energy regime at late times, spontaneous symmetry breaking could take place deriving the system to the actual minimum. In this case, the massless field would acquire mass (proportional to the density of massive neutrino matter) in the true ground state. If it happens, the mass of the field would naturally get connected to the energy density of massive neutrino matter.\\  
In order to implement the aforesaid idea, we propose to  consider the following action  
\begin{eqnarray}
\label{eq:action}
\S &=& \int  \d^4 x\sqrt{g}\left[\frac{\Mpl^2}{2}R-\frac{1}{2} \partial^\mu\phi\partial_\mu\phi-V(\phi)\right]   \nn \\ && + \mathcal{S}_\m+\mathcal{S}_\r+\S_\nu(\phi;\Psi) \, ,
\end{eqnarray}
$\S_\m$ and $\S_\r$ are the actions for matter and radiation
respectively and $\S_\nu$ is the action for massive neutrinos
(subscripts {\it m, r} and $\nu$  designate matter, radiation and
massive neutrinos respectively), see Ref.\cite{sajad} on related theme. We have considered a separate
action for massive neutrinos since it behaves as radiation at very
high red-shifts and becomes non-relativistic only in the recent past.
Also, we have considered a non-minimal coupling between the massive
neutrinos and scalar field which implies that neutrino masses acquire field dependence,
\begin{eqnarray}
\mathcal{L}_\nu=i\bar{\Psi}\gamma^\mu\partial_\mu\Psi-m_\nu(\phi) \bar{\Psi}\Psi
\label{mass}
\label{Lnu}
\end{eqnarray}
where $m_\nu(\phi)=A(\phi)m_{\nu0}$\footnote{Here $m_{\nu0}\equiv m_\nu(\phi=0)$}. In what follows, we shall make a specific choice for $A(\phi)$.

In flat Friedmann-Lema$\hat{\rm i}$tre-Robertson-Walker (FLRW) background the Friedmann equations,
corresponding to the action~(\ref{eq:action}) are
\begin{eqnarray}
 3H^2 \Mpl^2 &=&\rho_\m+\rho_\r+\rho_\nu+\frac{1}{2}\dot\phi^2+V(\phi) \, ,
 \label{eq:Fried_scal_inf}\\
 \left(2\dot H+3H^2\right)\Mpl^2 &=&-\frac{1}{3}\rho_r-p_\nu-\frac{1}{2}\dot\phi^2+V(\phi) \,, 
\end{eqnarray}
where $\rho$'s and $p$'s are density and pressure  respectively.

In presence of coupling, scalar field dynamics is governed by
the following evolution equations,
\begin{equation}
 \Box \phi=V_{,\phi}-\frac{A_{,\phi}}{A}T_\nu \, ,
\end{equation}
which in flat FLRW background gives
\begin{equation}
 \ddot\phi+3H\dot\phi=-V_{,\phi}+\frac{A_{,\phi}}{A}T_\nu .
 \label{eq:eom_scal}
\end{equation}

The coupling also reflects on the continuity equation for the massive neutrino matter,  
\begin{equation}
 \dot\rho_\nu+3H(\rho_\nu+p_\nu)=\frac{A_{,\phi}}{A}\dot\phi(\rho_\nu-3p_\nu) \, .
 \label{eq:cont_nu_1}
\end{equation}
In terms of $\hat\rho_\nu=A \rho_\nu $ where $\hat\rho_\nu$ is
independent of $\phi$,  the continuity
equation~(\ref{eq:cont_nu_1}), takes the standard form as  massive
neutrinos become non-relativistic
and Eq.~(\ref{eq:eom_scal}) becomes
\begin{equation}
 \ddot\phi+3H\dot\phi=-V_{,\phi}-A_{,\phi}\hat\rho_\nu \, .
 \label{eq:eom_scal_1}
\end{equation}
In what follows, we shall consider massless scalar field with $\lambda \phi^4$ potential and make a convenient choice for the coupling function $A(\phi)$ adhering to $Z_2$ symmetry,
\begin{eqnarray}
\label{conf}
 V(\phi)=\frac{\lam}{4}\phi^4;~ 
 A(\phi)=1-\frac{\alpha \phi^2}{2 \Mpl^2}
 \label{eq:pot}
\end{eqnarray}
where $\alpha$ is a constant to be fixed using observational constraints or some additional requirement\footnote{Actually, $\alpha$ in (\ref{conf}) defines a mass scale, $M\equiv \alpha^{-1}\Mpl $.}.
From the right hand side of Eq.~(\ref{eq:eom_scal_1}), we then obtain an
effective potential (up to an irrelevant constant),
\begin{equation}
 V_{\rm eff}=-\frac{\alpha \hat \rho_\nu}{2\Mpl^2}\phi^2 +\frac{\lambda}{4}\phi^4\, .
 \label{eq:pot_eff}
\end{equation}
Let us note that the assumed field dependence of neutrino mass led to the coupling of $\phi$ to massive neutrino matter (expressed by the first term in equation~(\ref{eq:pot_eff})).  The coupling could be motivated by conformal transformation from Jordan to Einstein frame,
\begin{equation}
\mathcal{S}=\int{ d^4x\sqrt{-g}\left[ \frac{\Mpl^2}{2}R-\frac{1}{2}(\nabla\phi)^2-V(\phi)  \right]}+\mathcal{S}_m+\mathcal{S}_r +\mathcal{S}(A^2(\phi)g_{\mu\nu}, \Psi_\nu)    
\label{E}
\end{equation}
where $g_{\mu\nu}$ is Einstein frame metric and $\Psi_\nu$ stands for (neutrino) matter field; the standard matter (cold dark matter, radiation ) is supposed to be minimally coupled. Variation of the action $\mathcal{S}$ with respect to $\phi$ gives rise to the field equation with effective potential (\ref{eq:pot_eff}) such that $\phi$ couples to neutrino matter through its matter density. 
In this case, no assumption is required  related to microscopic interaction of $\phi$ with matter field such as present in (\ref{mass}). If one deals with the interaction of field with neutrino matter through its matter density, the kind of average description cosmology is required. However, the mass dependence of neutrino mass is implied by the conformal transformation though not explicitly seen. Indeed, 
the  energy momentum tensor of neutrino matter transforms under conformal transformation giving rise to field dependence of neutrino mass (see, Ref.\cite{samrev} for details) introduced in (\ref{mass}) by hand.

Let us now examine the implications of coupling for dynamics. Thanks to the presence of coupling, the effective potential~(\ref{eq:pot_eff}) has minima at\footnote{where we introduced the notation $\hat\rho_{\nu(min)}\equiv \hat\rho_{\nu}$ for keeping track of the minimum.}
\begin{equation}
 \phi_{\rm min}=\pm\sqrt{\frac{\alpha}{\lambda}\frac{\hat\rho_{\nu(min)}}{\Mpl^2}} \, .
\end{equation}

\begin{figure}[ht]
\centering
\includegraphics[scale=.4]{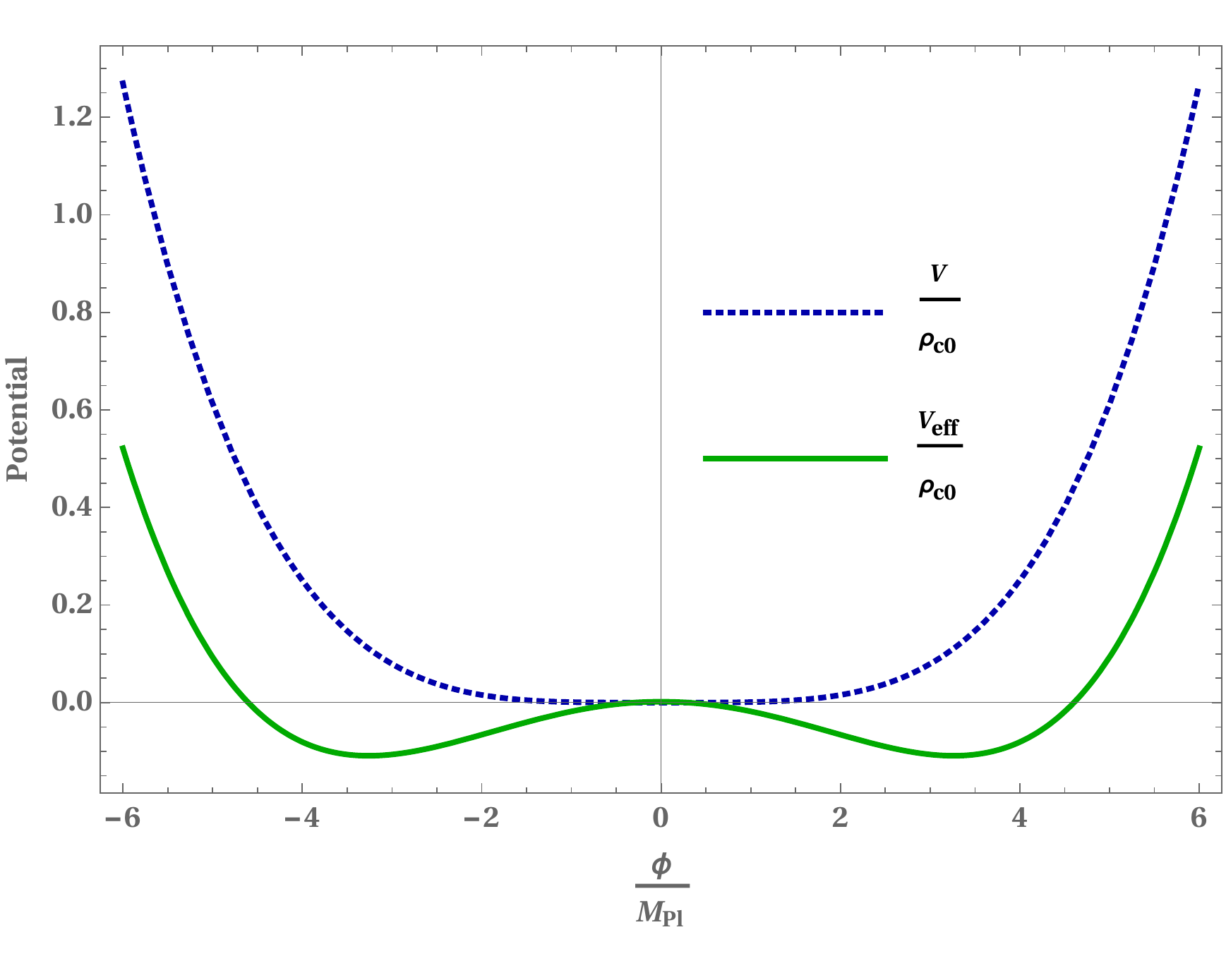}
\caption{Schematic plot of effective potential $V_{\rm
eff}/\rho_{c0}$  along with the original $\lambda
\phi^4$ potential  versus $\phi/\Mpl$ which shows that
symmetry breaking takes place in the system. As massive neutrinos
turn non-relativistic, transition to actual ground state takes
place.} 
\label{fig:Veff_analyt}
\end{figure}

The mass of the field around the true minimum and $V_{\rm min}$ is
given by,
\begin{equation}
m^2_{\rm eff}=2\alpha \frac{ \hat\rho_{\nu(min)}}{\Mpl^2};~V_{\rm
min}=-\frac{\alpha^2 \hat\rho^2_{\nu(min)}}{4\lambda \Mpl^4}.
\end{equation}
 Assuming that the true vacuum occurs around the present epoch, we estimate the mass of the field after spontaneous symmetry breaking,
\begin{equation}
m^2_{\rm eff}=2\alpha \frac{\hat\rho_{\nu(\rm min)}}{\Mpl^2}\sim H_0^2 \to
\hat\rho_{\nu(min)}\simeq \frac{\rho_{c0}}{6\alpha} \, ,
\end{equation}
where $\rho_{c0}$ is the present critical density. The above relation implies that $\alpha \simeq (6 \Omega_\nu^{(0)})^{-1} $, where $\Omega_\nu^{(0)}=0.02 (m_\nu/1eV)$. Clearly,
$\hat\rho_{\nu(\rm min)}$  gets connected to dark energy,
\begin{equation}
    \hat\rho_{\nu(\rm min)}=\left(\frac{\Omega^{(0)}_{\nu}}{\Omega^{(0)}_{DE}}\right)  \rho_{DE}^{(0)}
\end{equation}
where the present day fractional energy density of neutrino matter, $\Omega^{(0)}_{\nu}$, is parametrized through neutrino mass .
As for the coupling
 $\lam$, it can be estimated using  expression of the effective potential, $V_{\rm eff}^{\rm min}\simeq \Omega_{DE}^{(0)}\rho_{\rm c0}$, namely,
\begin{equation}
|V_{\rm eff}^{\rm min}|=\frac{\alpha^2 \hat\rho^2_{\nu(\rm min)}}{4\lambda \Mpl^4}\to
\lambda \simeq \frac{\rho_{c0}}{36 \Mpl^4}\left(\frac{1}{4\Omega_{DE}^{(0)}} \right)\simeq 10^{-123}.
\end{equation}
 The incredibly small
numerical value of self coupling might become problematic if we want to couple the field to any other matter field. We shall come back to this important point later where we suggest a way to circumvent the said problem.
Raising the potential by constant, we have\footnote{The effective potential (\ref{eq:pot_eff}) is defined up to an irrelevant constant.},
\begin{equation}
 V_{\rm eff}=-\frac{\alpha \hat \rho_{\nu(min)}}{2\Mpl^2}\phi^2 +\frac{\lambda}{4}\phi^4+2|V_{\rm eff}^{\rm min}|\, .
 \label{potf}
\end{equation}
We note that in (\ref{potf}), all the three terms at minimum are of
the same order.
 Using the values of $\lam$ and $\hat\rho_{\nu{(min)}}$ we plot the
effective potential (\ref{eq:pot_eff}) in
Fig.~\ref{fig:Veff_analyt} which clearly shows the emergence of actual ground state around the present epoch.

Let us now confirm that field indeed rolls slowly around the minimum. In the small
neighborhood of the minimum, the effective potential can be approximated by the following simple expression,
\begin{equation}
\label{ve} V_{\rm eff}\simeq \Omega_{DE}^{(0)}\rho_{\rm
c0}+\frac{\rho_{c0}}{6\Mpl^2} \delta\phi^2
\end{equation}
where we have kept lowest order term in $\delta \phi$. The validity
of the required slow roll around the minimum can be checked using the
estimate,
\begin{equation}
\epsilon\equiv
-\frac{\dot{H}}{H^2}=\frac{\Mpl^2}{2}\left(\frac{V'_{\rm eff}}{V_{\rm
eff}}\right)^2=\frac{3}{2}(1+\omega_\phi),
\end{equation}
which for the observed valued of equation of state parameter of dark
energy gives, $\delta \phi\lesssim 0.7 \Mpl$ such that the
approximation used in (\ref{ve}) is valid.
\begin{figure}[ht]
 \centering
\includegraphics[scale=.4]{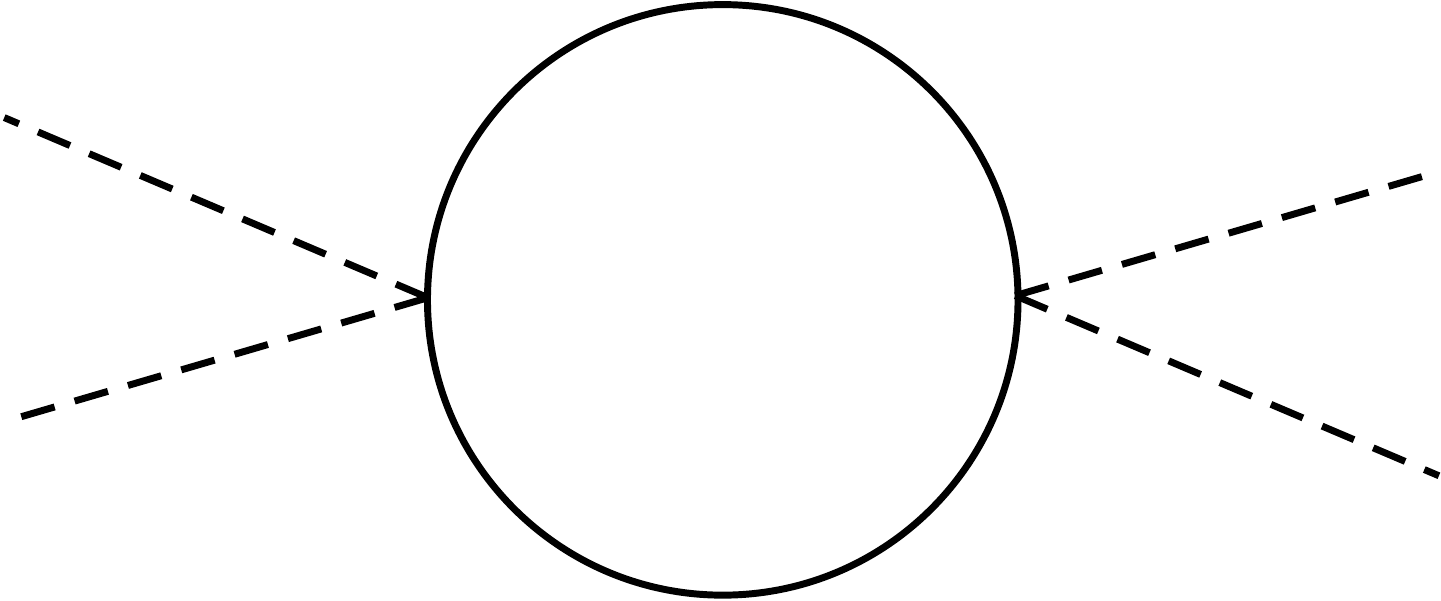}
\caption{One loop diagram based upon the effective interaction: $ \frac{g}{M}\phi^2\bar{\Psi}\Psi $ where $M$ is a cut off or mass scale. The dashed lines correspond to $\phi$ field and the closed loop is a fermionic loop. This diagram generates $\lambda \phi^4$ interaction. }
\label{circl}
\end{figure}

\begin{figure}[ht]
 \centering
\includegraphics[scale=.4]{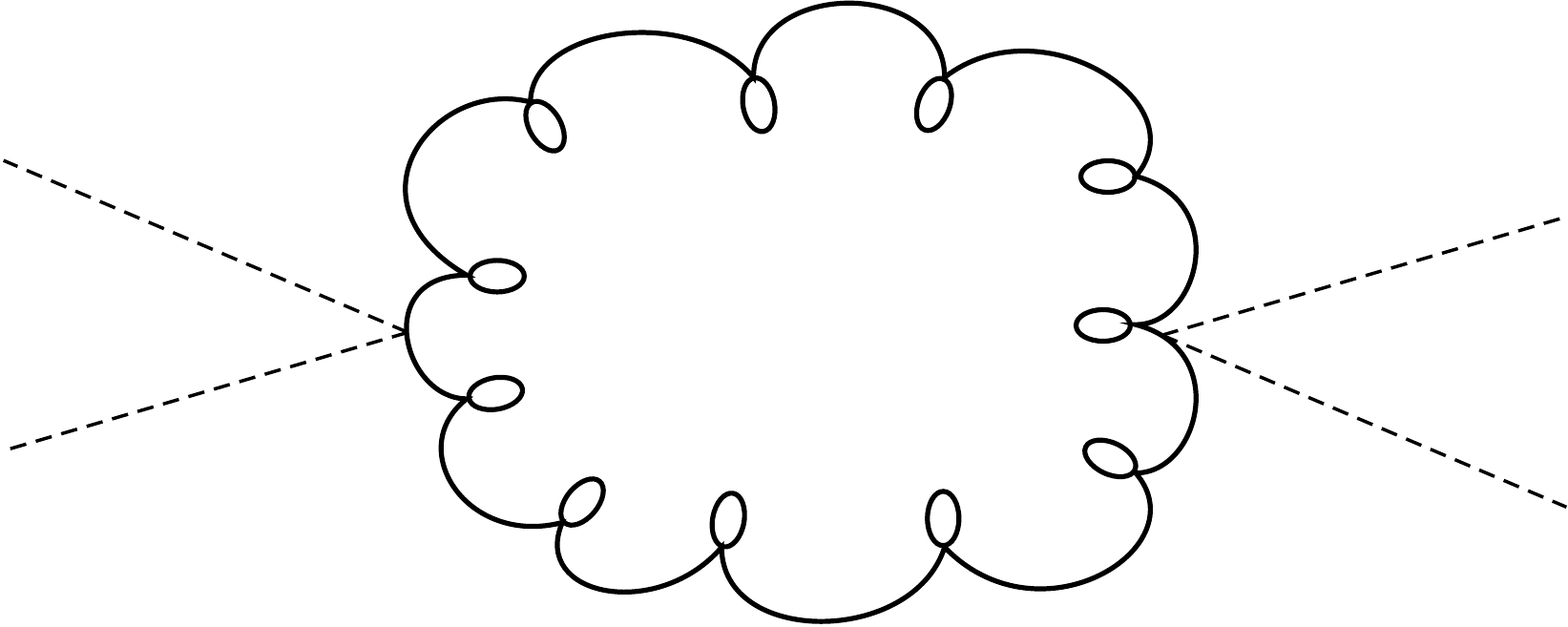}
\caption{One loop diagram  generated by $\phi$ interaction with graviton. The curly closed loop is a graviton loop. The diagram gives rise to a correction to the coupling constant $\lambda$ which is suppressed by $\Mpl$.}
\label{graviton}
\end{figure}
We should hereby point out  to a serious problem due to a small value of the self coupling $\lambda$. Its incredibly small numerical value would be difficult to protect under radiative corrections in case $\phi$ interacts with the matter field. {  Before getting into the discussion to follow, we should  emphasize that $\phi$ is not the part of standard model$-$ it does not carry standard model quantum numbers, it is a singlet. A slowly rolling scalar field with mass of the order of $10^{-33}$ eV is necessary to reconcile with the late time acceleration of Universe thereby $\phi$ can not be associated with Higgs field. In the standard model , the Higgs field has SU(2) doublet structure, part of which provides the longitudinal components to gauge fields($Z,W^{\pm}$) making them massive. The remaining neutral component H is the physical Higgs particle. In our case,
the interaction of $\phi$ with fermions, gauge fields and gravitons are effective interactions, to be considered purely in the phenomenological setting. However, some of these interactions might be inspired by  standard model.}

\begin{figure}[ht]
 \centering
\includegraphics[scale=.4]{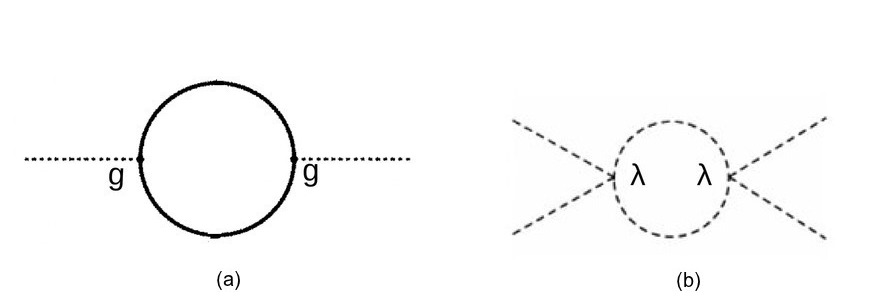}
\caption{Fermion and boson loops generated from interactions$-$ $g_F\phi \bar{\Psi}\Psi$(a) and $\lambda \phi^4$(b). Dashed lines correspond to $\phi$ where as solid line corresponds to fermion field. These loops give rise to quadratic and logarithmic divergences respectively.}
\label{fermion}
\end{figure}

\begin{figure}[ht]
 \centering
\includegraphics[scale=.4]{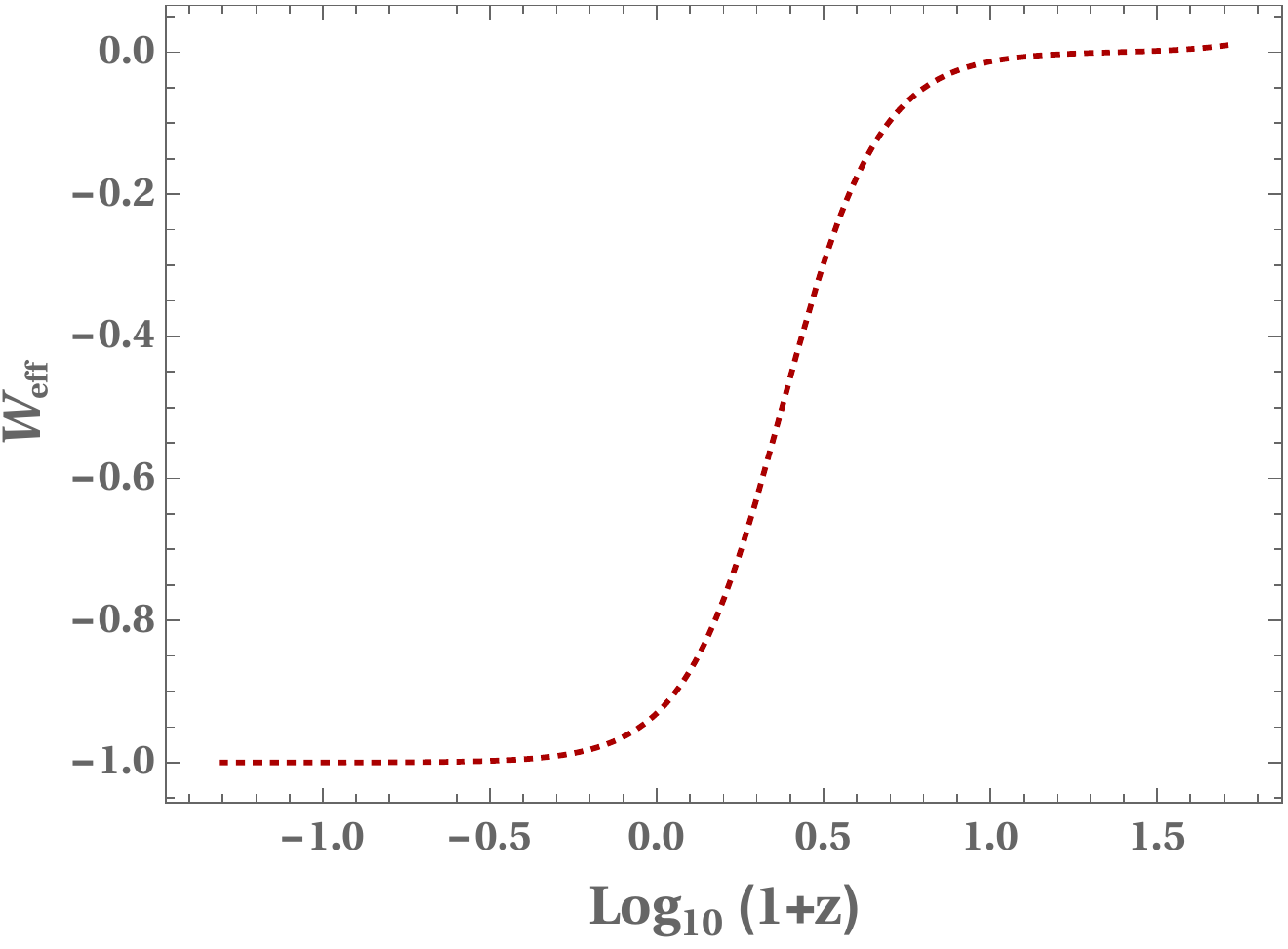}
\caption{Plot of effective equation of state $w_{eff}$ obtained from numerical integration of evolution equations versus the red-shift(for $\lambda =10^{-123}$ and the $m_\nu=10^{-2}eV$). Plot confirms our analytical assertion about the onset of de-Sitter phase.}
\label{weff}
\end{figure}

We could try to generate $\lambda \phi^4 $ term in the Lagrangian by one loop correction from the following effective interaction, 
\begin{eqnarray}
\label{dim5}
\mathcal{L}_{int}=\frac{g}{M}\phi^2 \bar{\Psi}\Psi
\end{eqnarray}
where $g$ is a dimensionless coupling. Since $\phi^2 \bar{\Psi}\Psi $ is a mass dimension five operator, we need a mass scale or cut off in the denominator in (\ref{dim5}). One loop diagram (Fig.\ref{circl}) is quadratic divergent and generates $\phi^4$ interaction with the self coupling given by,
\begin{eqnarray}
\lambda \sim g^2 \frac{\Lambda^2_c}{M^2}\ln{\Lambda^2_c}
\end{eqnarray}
 where $\Lambda_c$ is UV cut off. For a generic cut off, $ {\Lambda^2_c}/{M^2}\sim 1$, thereby $\lambda\sim g^2$. However, we should further worry about radiative corrections to $g$ or equivalently to $\lambda$. The corrections might come from graviton loops~\cite{hooft,dongu} as well as from gauge field loops shown in Fig.\ref{graviton}. Let us first examine the contribution from graviton loop, Fig.(\ref{graviton}). The two scalar-two  graviton vertex is given by,
\begin{eqnarray}
&&\tau_{2-2}^{\eta\lambda\rho\sigma}(p,p',m)=\frac{4i}{\Mpl^2}\big[I^{\eta \lambda \alpha \delta}I^{\rho\sigma\beta\delta}\nonumber \\
&&-\frac{1}{4}\left(\eta^{\eta \lambda}I^{\rho\sigma\alpha\beta}+\eta^{\rho\sigma}I^{\eta\lambda\alpha\beta}\right)(p_\alpha  p'_\beta+p_\beta p'_\alpha)\nonumber\\
&&-\frac{1}{2}\left(I^{\eta\lambda\rho\sigma}-\frac {1}{2}\eta^{\eta\lambda}\eta^{\rho\sigma}\right)(p.p'-m^2)\Big],\nonumber\\
&& I_{\alpha\beta\rho\sigma}=\frac{1}{2}\left(\eta_{\alpha\rho}\eta_{\beta \sigma}+
\eta_{\alpha\sigma}\eta_{\beta\rho}\right)
\end{eqnarray}
where $p$ and $p'$ designate four momenta carried by $\phi$ lines and $m$ the mass of the scalar field. The diagram in Fig.\ref{graviton} gives rise to logarithmic divergence and its contribution is suppressed by the powers of $\Mpl$. Thus the self coupling generated by the diagram in Fig.\ref{circl} is protected under radiative corrections coming from graviton loops. Unfortunately, self coupling receives large corrections from $\phi$ interaction with gauge fields. For instance, one loop correction  generated from $\phi$ interaction with gauge field\footnote{The coupling of $\phi$ to gravitons or gauge fields are purely phenomenological; in general the field $\phi$ is not the part of standard model of particle physics.},
\begin{eqnarray}
\mathcal{L}_{int}=g_G^2\phi^2W_{\mu} W^\mu(Z_\mu Z^\mu)  \to 
\delta \lambda\sim g_G^4 \ln{\Lambda^2_c}
\end{eqnarray}
 is pretty large given that $\lambda \sim 10^{-123}$ assuming that $g_G$ is of the order of electroweak coupling. 
A comment on the radiative correction to the mass of $\phi$ with bare mass zero  in (\ref{eq:action}) is in order. If we admit its coupling to matter fields $g_F\phi \bar{\Psi}\Psi$, the one loop correction (Fig. \ref{fermion}(a)) from the highest mass fermion circulating in the loop
is prominent,
\begin{equation}
 \delta m^2\sim  g_F^2 m_s^2\ln(\mu^2_D/\Lambda^2_c)   
\end{equation}
where $m_s$ is the highest mass of fermion available, for instance, mass of t quark in the standard model; $\mu_D$ is an arbitrary mass scale in method of dimensional regularization.
This correction is huge otherwise coupling is negligibly small. In fact such a large correction is the artifact of quadratic divergence $\hat{\rm a}$  {\it  la} 't Hooft  naturalness. As for the one loop correction to $\lambda$ due to self coupling of $\phi$ (see Fig.\ref{fermion}(b)), $\delta \lambda \sim \lambda^2 \ln \Lambda^2_c$ $-$ result of logarithmic divergence which does not disturb the small value of coupling set by observation\footnote{We thank R. Kaul for clarifications on related issues}.
Lets us  also mention that in symmetron scenario\cite{Hinterbichler:2010es}, the starting point
was scalar field with Higgs like potential with wrong sign of mass
scale $\mu$,
\begin{equation}
V_{\rm
eff}=\frac{1}{2}\left(\frac{\rho}{M^2}-\mu^2\right)\phi^2+\frac{\lambda}{4}\phi^4
\end{equation}
where $\rho$ is standard matter density. In high density regime,
$Z_2$ symmetry is exact. Symmetry breaking takes place when matter
density falls around the critical density in the Universe. Local
physics then imposes an stringent constraint on the mass scale $M$
such that the mass of the field at late times turns out about $10^4
H_0$\cite{Hinterbichler:2010es}. As a result, field rolls fast around the true minimum and
keeps overshooting it for a long time. Symmetry breaking is cleverly
managed in the high density regime but late time acceleration is beyond the scope of this
scenario$-$  "No Go" theorem. In our proposal, we circumvented problem by bringing the
coupling with massive neutrino matter which shows up at late stages
at large scales and is free from local gravity constraints. Let us point out that the true minimum of the effective potential occurs far away from the origin which might look problematic at the onset. Indeed, using the observed values of density parameters, we have
\begin{eqnarray}
\phi_{min}=\left( \frac{4\Omega^{(0)}_{DE}}{\alpha\Omega_\nu^{(0)} }\right)^{1/2}\Mpl\simeq 4 {\Mpl}
\end{eqnarray}
irrespective of the numerical value of $\alpha$, where we used the constraint, $\alpha=(6\Omega_\nu^{(0)})^{-1}$ which arises from the requirement that $\phi$ rolls slowly around the minimum. Actually, $\alpha$ defines mass scale in Eq.(\ref{conf}), namely, $M\equiv \alpha^{-1/2} \Mpl$ which is suitable to slow roll\footnote{Let us note that  the observed values, $0,0022\lesssim\Omega^{(0)}_\nu \lesssim 013$ correspond to $1.3 \lesssim \alpha \lesssim 76$ or equivalently,   $0.12 \lesssim M/\Mpl \lesssim 0.8$.} demonstrated by analytical estimates. We have have numerically integrated the equations of motion to confirm our analytical results; see Fig.\ref{weff} which shows the onset of de-Sitter regime.

Let us note that before the field could roll to the minimum, $A(\phi)$ turns negative when $\phi=\phi_0\simeq 0.5 \Mpl$ turning the neutrino masses negative which might look problematic. 
Interestingly, changing $\alpha$ does not help,
\begin{eqnarray}
\frac{\phi_{min}}{\phi_0}=\sqrt{2} \left(\frac{ \Omega_{DE}}{\Omega_\nu^{(0)}}  \right )^{1/2};~~A(\phi_0)=0 \,,
\end{eqnarray}
which means that $A(\phi)$ turns negative much before minimum is reached ($\Omega_\nu^{(0)}<<\Omega_{DE}$) irrespective of the choice of mass scale. 
Let us note that all the physical quantifies involving neutrino masses depend upon their even powers rather than masses themselves. {The negative mass or negative energy states can be interpreted  as anti-particle states with positive energy $\hat{\rm a}$  {\it  la}  Dirac-Feynman. Thus changing sign of $m$ in Dirac equation may be thought as renaming particles by anti-particles and vice-versa. From the field theory perspective\footnote{Field theoretic framework does not include negative energy states.},  let  us emphasize that replacing mass $m$ in Dirac Lagrangian by $-m$ does not change the underlying physics. Indeed, Dirac equation can be written as set of two first order coupled differential equations for Left and Right components which are independent fields such that  coupling between them is provided by the mass $m$ term. Replacing $m$ by $-m$ amounts to multiplying the Left (Right) component of the Dirac field by $-1$; obviously redefining the field  does not give rise to new  physics.}

In view of the aforesaid, it is clear that in our setting, we  capture late time cosmic acceleration which was excluded in the original symmetron model by the requirement of local gravity constraints. 
Clearly, our scenario,  allows to realize symmetry breaking in the low density regime giving rise to late time cosmic acceleration.

Before concluding our findings, let us compare our scenario with the standard framework with  field dependent neutrino masses. To this effect, we consider a scalar field with a steep run away type of potential such that the scaling behaviour could be realized. We further assume that neutrino matter is non-minimally coupled such that, $S(A^2(\phi) g_{\mu\nu},\Psi_\nu)$ with $g_{\mu\nu}$, $\Psi_\nu$ being the Einstein frame metric and the matter field for massive neutrino matter respectively; radiation and standard matter (DM plus baryonic matter) are taken to be minimally coupled.  Further choosing the coupling, $A(\phi)\sim \exp{(\gamma\phi/M_{Pl})}, \gamma>0$, we have \cite{H1},
\begin{equation}
\label{rhol}
V_{eff}=V(\phi)+\rho_{\nu0}e^{\gamma( \phi-\phi_0)}/M_{Pl}
\end{equation}
where $\phi_0$ refers to the field value today. In the second term in (\ref{rhol}), $\rho_{\nu 0}$ comes from the trace of neutrino matter which builds up only at late times as neutrino turn non-relativistic. In the nutshell, the effective potential has a minimum mimicking de Sitter like behaviour at late stages\footnote{With our choice of the conformal factor which is dictated by the requirement to trigger minimum in the run away potential, neutrino mass grows exponentially. It is not desirable to have non-minimal coupling to standard matter as the latter would induce minimum in the field potential soon after matter domination is established spoiling the matter phase. }.  However, we have preferred to work
in the physical Jordan frame assuming the neutrino masses there to be field dependent. 
In model under consideration, we have  scalar field with $\lambda \phi^4$ potential which has minimum at $\phi=0$, the choice of $A(\phi)$ in the effective potential is such that tachyonic instability builds up at late states leading to true ground state around which field would evolve slowly giving rise to late time acceleration. Thus our choice here, unlike the standard  mass varying neutrino framework, is associated with the requirement of spontaneous symmetry breaking in the low density regime capable of mimicking dark energy like behaviour at late times. The latter was ruled out in symmetron model due to local gravity constraints one has to adhere to in case of coupling to standard matter; neutrino matter is free from such constraints.

In this paper, we have attempted to realize the late time phase transition in Universe using massless $\lambda \phi^4$ theory, non-minimally coupled to massive neutrino matter. In this framework,  as massive neutrinos start getting non-relativistic, coupling to the field gradually builds up dynamically giving rise to tachyonic instability in the system. As a result, the transition to a actual ground state is realized~\footnote{ We should note that the phase transition under consideration is a slow process, it commences when $z\sim z_{eq}$ and is completed as massive neutrino matter settles to cold dark matter like state around the present epoch.}.
Consequently, the field acquires a non-zero mass in the true vacuum and the field energy density in the potential minimum is given by massive neutrino matter  density which could naturally be connected to dark energy. We have shown that field rolls slowly around the actual minimum and can give rise to late-time cosmic acceleration in given parameter space. In this case, however, the minimum is realized far away from the origin such that before the field could roll to the minimum, the conformal coupling becomes negative turning the neutrino masses negative. The negative energy states can be understood as anti-particle states with positive energy using interpretation $\hat{\rm a}$  {\it  la} Dirac-Feynman. The latter amounts to renaming neutrinos by anti-neutrinos and vice-versa which does not lead to new physics; the latter can also be understood in the field theoretic framework.

The distinguished and generic feature of our proposal is related to the physical process of turning massive neutrinos to non-relativistic at late times which gives rise to the breaking of $Z_2$ symmetry in the low density regime. However, there is a price to be paid, namely, the coupling $\lambda$ is negligibly small similar to the case of symmetron~\cite{Hinterbichler:2010es}. The latter would exclude interaction of $\phi$ with any matter field, i.e., an incredible fine-tuning for the coupling with matter fields, otherwise, $\lambda$ would receive large quantum corrections. In order to circumvent this problem, we proposed to generate $\lambda \phi^4$ interaction from one loop correction induced by mass dimension five operators, $ \phi^2 \bar{\Psi}\Psi$. In this case, the self-coupling is stable under radiative corrections from graviton loops. However, the interaction of $\phi$ with gauge fields, if it exists, might\footnote{ We emphasize that the field $\phi$ in our model is not the part of standard model of particle physics and its interactions with other fields, in particular with gauge fields and gravitons, are purely phenomenological in character. },  generate large corrections to self-coupling. 
A remark related to the coupling of $\phi$ with gauge fields, gravitational field and matter fields responsible for the radiative corrections to the mass of $\phi$ and self coupling $\lambda$ is in order. The $\phi$ dependence of neutrino mass in (\ref{mass}) leads to the coupling of the field to neutrino matter through its matter density in~(\ref{eq:pot_eff}).
After spontaneous symmetry breaking, the system settles to its true ground state at late times and neutrino mass settles to its present (constant) value thereby there is no coupling of $\phi$ to matter fields $\Psi $ due to  mass term in (\ref{mass}) in the true ground state {\it a la} de-Sitter. The couplings of the field with gauge fields/gravitational field and matter fields are of phenomenological nature with $\Psi$ being any matter field not necessarily neutrino field.

Finally, we have  demonstrated  that the "No Go" result in the symmetron model associated with local gravity constraints is evaded by bringing in the
coupling of massive neutrino matter to the scalar field.
Interestingly, in this process, the
dimensionless density parameter of dark energy $\Omega^{(0)}_{DE}$ gets naturally connected to $\Omega^{(0)}_{\nu}$, the only physical energy scale available in the late Universe. It would be interesting to study perturbations  and investigate 
their implication for matter power spectrum in the scenario under consideration; we deffer the same to our future investigations.

\section*{ACKNOWLEDGMENTS}
The authors are  thankful to A. Starobinsky, R. Kaul, R. Adhikari, V. Soni, Wali Hossain, N. Jaman and C. Lee for useful discussions. We are indebted to  Mohseni Sadjadi for his detailed comments on the draft.

\end{document}